\documentclass[10pt, twocolumn,amsmath,amssymb,aps]{revtex4}

\usepackage[dvipdfmx]{graphicx,color}
\usepackage{dcolumn}
\usepackage{bm}
\usepackage{color}
\usepackage{array}
\newcommand{\be}{\begin{equation}}
\newcommand{\bea}{\begin{eqnarray}}
\newcommand{\ee}{\end{equation}}
\newcommand{\eea}{\end{eqnarray}}

\begin{document}

\title{Dynamics of fermions in an amplitude modulated lattice}
\author{Tomotake Yamakoshi$^{1}$, \footnote{corresponding author: shin.watanabe@uec.ac.jp}Shinichi Watanabe$^{1}$, Shun Ohgoda$^{1}$,  and Alexander P. Itin$^{2,3}$}

\affiliation{$^1$University of Electro-Communications, Chofu-shi, Tokyo 182-8585, Japan, \\
$^2$Radboud University, Institute for Molecules and Materials (IMM), Nijmegen 6525 AJ, The Netherlands, \\
$^3$Space Research Institute, Russian Academy of Sciences, Moscow 117997, Russia }
\begin{abstract}
We  study  dynamics of fermions loaded in an optical lattice with a superimposed parabolic trap potential.  In the recent Hamburg experiments  [J.Heinze {\it et.al.}, Phys. Rev. Lett. {\bf 110}, 085302 (2013)] on quantum simulation of photoconductivity, a modulation pulse on the optical lattice transferred part of the population of the lowest band to an excited band, leaving a hole in the particle distribution  of the lowest band. Subsequent intricate dynamics of both excited particles and holes can be explained  by a semiclassical approach based on the evolution of Wigner function. Here we provide a more detailed analysis of the dynamics taking into account the dimensionality  of the system and  finite temperature effects, aiming at reproducing experimental results on longer timescales.  A semiclassical wave packet is constructed more accurately than in the previous theory. 
As a result,  semiclassical dynamics indeed reproduces experimental data and full quantum numerical calculations with much better accuracy.  In particular, fascinating phenomenon of collapse and revival of holes is investigated in a more detail. 
We presume the experimental setup can be used for deeper exploration of nonlinear waves in fermionic gases. 
\end{abstract}

\maketitle

\section{introduction}
\label{sect:Introduction}

Ultracold atoms and molecules  in optical lattices  may provide a path to construction of quantum simulators (well-controllable systems whose dynamics allow to understand essential features of more complicated condensed matter systems \cite{Feynman, Lewenstein}).
Recent examples include numerous simulations of Hubbard-type models in optical lattices \cite{OL1,
SIMU, GAUGE, LBO, GRAVM, LZZ}, in particular the simulation of photoconductivity in \cite{Heinze}.
At the same time,  any such simulator,  once created, possesses its own intricate dynamics (in some sense, trying to live its own life, and not to be a simplistic version of another system).  Such type of behavior indeed showed up in the Hamburg experiment on quantum simulation of photoconductivity  \cite{Heinze} recently,   giving rise to  unexpectedly rich dynamics even in the case of single-component (non-interacting) fermions.
We note that dynamics of many-particle fermionic systems is very interesting even in the absence of interactions, and one can indeed study a variety of effects including quantum-classical correspondence, nonlinear waves, {\it etc}.   For example, in \cite{Protopopov} dynamics of a density pulse induced by a local quench in a one-dimensional electron system was studied. The spectral curvature led to an ``overturn" (population inversion) of the wave,  after which the density profile developed strong oscillations. Straightforward realization of such a setup in an optical lattice would require a hugely long optical lattice with thousands of sites.  
However, one may try to realize this phenomenon in a more compact parabolic optical lattice, in which case the effects found in \cite{Heinze} and studied in a more detail here are relevant.
Another interesting topic is the control of matter waves by high-frequency driving. Not only the band dispersion can be modified by driving \cite{Holthaus,Eckardt,Korsch}, but it is also possible
to alter {\it e.g.} effective interactions in a system of interacting atoms (like exchange coupling between effective spins, {\it etc.} \cite{MI,Johan}). In the context of the photoconductivity
simulation experiment \cite{Heinze}, one may engineer the excited matter wave packet by varying the length and strength of the modulation pulse.

In this paper, we focus on the dynamics of non-interacting fermions in the amplitude modulated lattices.
Though examined {\it experimentally}  by the Hamburg group\cite{Heinze}, it is not {\it theoretically} investigated in sufficient depth. 
Here we provide some exact numerical results and a semiclassical description.
Especially, we reexamine the hole dynamics that prevails after the holes are created by the amplitude modulation. 
The temperature effect and the anharmonicity of the trap are also taken into account in our numerical simulation. 
We find that the hole has intrinsically long coherence time.

In the next Section, we give some additional background and describe our system and methods. 
Section III contains comparison of experimental dynamics with 1D quantum and semiclassical simulations.
Section IV studies 3D dynamics, including effects of the temperature and anharmonicity of the trap. 
Section V contains conclusions.

\section{The system}
\label{sect:system}

We consider an optical lattice in a parabolic potential filled by spinless or spin-polarized fermions,  as in most experiments such as \cite{Heinze}.
Some  techniques and notations used in this paper are also available in the experimental work \cite{Heinze}, and  in a recent numerical study \cite{Yamakoshi} on single-particle dynamics. 
We use recoil energy $E_r=\hbar^2 k_{r}^2/2 m$ for the unit of energy, recoil momentum $k_{r}=2\pi/\lambda$ for the unit of (quasi-)momentum, lattice constant $a=2/\lambda$ for the unit of length and rescaled time $t=E_r t'/\hbar$ for the unit of time. Here $\hbar$, $\lambda$ and $m$ correspond to the Planck constant, wave length of the optical lattice, and mass of the particle, respectively. 
To reproduce the experiment numerically, we employ a more accurate numerical setup than in \cite{Heinze},  extending it  to 3D geometry, and partly taking into account temperature effects by preparing initial conditions correspondingly.  In accordance with the experimental procedure, we apply a modulation pulse to the lattice which excites part of the atomic population from the lowest to excited bands.  After the pulse, the system is left to evolve for a certain time, and then the band mapping procedure is fulfilled by ramping down the lattice.
The resulting particle momentum distribution gives insight into the dynamics of fermions after the excitation.  Importantly, we can also evaluate the corresponding semiclassical distribution more accurately  than previously\cite{Heinze}.

The 1D version of the system is described by the time-dependent Hamiltonian
$$H=-\frac{\hbar^2}{2m}\frac{\partial^2}{\partial x'^2}+ V_0\sin^2 (k_{r}x') [1+\epsilon_0(t') \cos(\omega ' t')] + \frac{1}{2}m\omega_0^2 x'^2$$ where $V_0$ is the height of the optical lattice, $\omega^\prime$ is the frequency of the amplitude modulation, $\epsilon_0(t')$ is the amplitude of the time modulation which, in the present paper, has a square shape in time, and $\omega_0$  determines the curvature of the trap potential. Rescaling the Hamiltonian, we get
\bea
H &=& -\frac{\partial^2}{\partial x^2}+s \sin^2 (x) [1+\epsilon_0 \cos(E_\omega t)]+ \nu x^2  \nonumber\\ &=& H_0 +s \sin^2 (x)\epsilon_0 \cos(E_\omega t),
\label{eq:re-ham}
\eea
where $x$, $s$, $E_\omega$ and $\nu$ denote $x=k_{r}x'$, $s=V_0/E_r$, $E_\omega=\hbar \omega ' / E_r$, $\nu=m\omega_0^2/2E_r k_{r}^2$, and $H_0=-\frac{\partial^2}{\partial x^2}+ s \sin^2 (x)+ \nu x^2$, respectively. The parameter $s$ gives the depth of the optical lattice in units of the recoil energy. As is usual in ultracold atom systems, this parameter can be easily controlled, and its typical value in the Hamburg experiment varied in the range of $s=2-20$.
 
Since the trap potential varies slowly as a function of $x$, one can use semiclassical approach based on dispersion of the unperturbed  periodic system without a trap \cite{Heinze}.
Let us recall the concept of quasimomentum in the uniform lattice system governed by $H_B=-\frac{\partial^2}{\partial x^2}+s \sin^2 (x)$.
The eigenstate $\phi_q^n$ of $H_B$ corresponding to energy $E_q^n$ is the Bloch state represented by
\begin{equation}
\phi_q^n(x)=e^{iqx} \sum_K C_B^n (K,q) e^{2iKx}, (n=0,1,2,...)
\label{eq:q-state}
\end{equation}
with suitable coefficients $C_B^n (K,q)$
where $n$, $q$ and $K \in \mathbb{Z}$ represent the band index, quasimomentum, and the corresponding reciprocal vector, respectively.  The set $\{\phi_q^n\}$ serves to label the eigenenergy states of $H$ in the static limit where $\epsilon_0(t^\prime)\equiv0$. Namely,
diagonalizing $H$ in the basis of the Bloch states, Eq.~(\ref{eq:q-state}) yields banded eigenenergies and coefficients labeled by quasimomentum $q$. We call the lowest energy band with the index $n=0$ the ground band hereafter.

The experiments of \cite{Heinze} were done both with interacting fermions composed of two-species and with non-interacting fermionic atoms of a single component.
Here we consider the latter case, non-interacting fermions. 
The initial wave function is obtained from the Slater determinant of single-particle wave functions which retains its form under the influence of time-dependent multi-particle  Hamiltonian  (being composed of time-dependent single particle wave functions).
This is because the Hamiltonian involves only single-particle operators. The expectation value of  the number operator
$$
\hat{n}=\sum_{j=1}^N \delta(x-x_j)
$$
with respect to the Slater determinant
\bea
\Psi(x_1,x_2,\cdots,x_N;t) =&\nonumber\\  \frac{1}{\sqrt{N!}}
\begin{vmatrix}
\psi_1(x_1;t)&\psi_2(x_1;t)&\cdots&\psi_N(x_1;t)\\
\psi_1(x_2;t)&\psi_2(x_2;t)&\cdots&\psi_N(x_2;t)\\
\vdots&\vdots&\vdots&\vdots\\
\psi_1(x_N;t)&\psi_2(x_N;t)&\cdots&\psi_N(x_N;t)
\end{vmatrix}
\label{eq:slater}
\eea
yields the same result as that of the Hartree  product $\Psi_{\rm Hartree}(x_1,x_2,\cdots,x_N;t)=\prod_{j=1}^N \psi_j(x_j;t)$ at all times, that is 
$$
\langle\Psi|\hat{n}|\Psi\rangle=\sum_{j=1}^N |\psi_j(x;t)|^2
$$
so that
$$
\int \langle\Psi|\hat{n}|\Psi\rangle \ dx=N,
$$
each single-particle state being normalized to unity.  This feature makes it convenient to study many-particle dynamics by preparing $N$ eigenstates
of a single-particle system and propagating them independently in a time-dependent potential, extracting overall density by mere summation of the single-particle densities.

\section{ Exact and semiclassical dynamics of fermions following a modulation pulse }
\label{sect:numerics}
Here we reproduce the experimental procedure of \cite{Heinze} numerically,  that is, first prepare the ground state of this fermionic system in the combined lattice and parabolic potential at temperature $T=0$,
excite it with a lattice amplitude modulation pulse, let it evolve, and make a band mapping by turning off the lattice potential and trap.

\begin{figure}[htbp]
 \begin{center}
 \includegraphics[width=7cm]{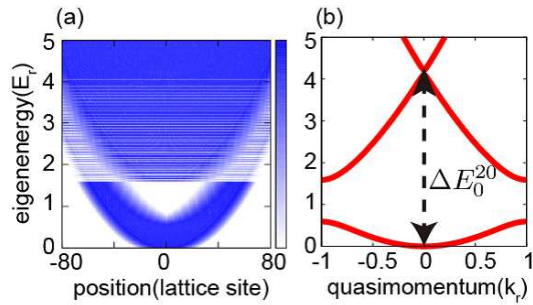}
 \end{center}
 \caption{(Color online) (a) Eigenstates of the Hamiltonian $H_0$ displayed at the corresponding energies as functions of position.
 The bluer, the higher probability density. The lowest eigenenergy is set to 0.
 The Fermi energy equals 0.97 when $N=81$ atoms are present. It is thus possible, for instance, to read off that the hole dynamics is constrained to the range of $(-60, 60)$ in position space.
 (b) Energy band structure of uniform lattice system $H_B$.
 To create a single hole at $q=0$ in the ground band, atoms are excited from ground to second band with excitation energy $E_\omega=\Delta E_0^{20}=4.223$.}
 \label{fig:spec}
\end{figure}

In Fig.~\ref{fig:spec}(a) we show the energy levels depicted as a function of position for the particular case of $s=2$ and $\nu=4.49\times 10^{-5}$.  
The figure displays the local atomic density $|\chi_j(x)|^2$ for each energy eigenstate of $H_0$ pertaining to eigenenergy $\epsilon_j$, using a darker shade for a higher density. 
To its right, Fig.~\ref{fig:spec}(b) shows, as a reference the energy bands in the absence of the harmonic confinement as a function of quasimomentum. 
In order to make a rigorous quantum simulation, we numerically solved the Time-Dependent Schr\"{o}dinger Equation(TDSE) with the initial Slater determinant, Eq.~(\ref{eq:slater}) and the total Hamiltonian $\sum_j H(x_j)$.
In this paper, we fix the depth of the optical lattice to $s=2$, the parabolic trap strength to $\nu=4.49\times 10^{-5}$, the modulation amplitude to $\epsilon_0 = 0.4$ and the pulse duration to 0.5ms, respectively, thus following the Hamburg experiment\cite{Heinze} closely.

\begin{figure}[htbp]
 \begin{center}
 \includegraphics[width=3.5cm]{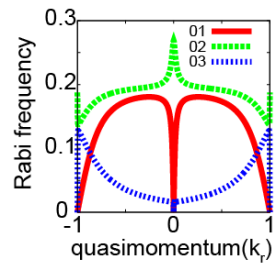}
 \end{center}
 \caption{Some Rabi frequencies for ground to first (red solid line), ground to second (green dashed line) and ground to third (blue dashed line) channels as functions of quasimomentum.
 See Ref.~\cite{Yamakoshi} for more detailed discussions.}
 \label{fig:Rabi}
\end{figure}

The parabolic trap varies gradually from site to site, so that we exploit the free band structure Fig.~\ref{fig:spec}(b) for labeling the bands in what follows.  On the other hand, the actual numerical calculations are done using the energy eigenstates of $H_0$.  The concept of Rabi interband oscillations  (oscillations between eigenstates belonging to different bands at the same quasimomentum) proved to be useful  in  ~\cite{Yamakoshi},  
 so that we show in Fig.~\ref{fig:Rabi} some Rabi frequencies $\Omega$ for the present system. 
 A resonant excitation occurs at such values of $q$, where the band energy difference matches $\hbar\omega^\prime$. 
The behavior of $\Omega$ over a  range of $q$ wider than in the bosonic case is thus relevant for fermions.
One observes that the frequency $\Omega$ for the 0th to the  second band transition has the largest overall values, and peaks near $q=0$. The Rabi frequency for the 0th to the first comes next, except that its value plummets exactly to zero at $q=0$. 

In the case of $q=0$ a single hole is created, while a pair of holes appear in the quasimomentum distribution when the modulation frequency and the inter-band energy difference match at some $q \not =0$.  
In the particular case of $s=2$, a single hole is formed approximately at $\omega^\prime = E_{q=0}^{0\rightarrow 2}/\hbar= 4.223E_r/\hbar$, so we fix the modulation energy to $E_\omega=4.223$. 

At the moment the amplitude modulation is turned off, a hole of a considerable depth is created in the ground band.
The subsequent time evolution of the matter wave packet has rich dynamical features. 

\begin{figure}[htbp]
 \begin{center}
 \includegraphics[width=8cm]{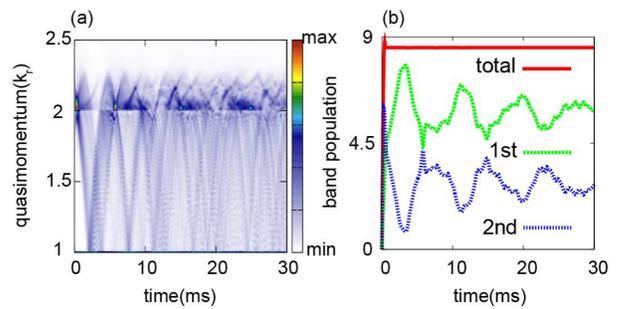}
 \end{center}
 \caption{(a) Quasimomentum distributions of the first and second bands. 
 Number of atoms $N=81$ corresponds roughly to the cubic root of $N_{3d}=10^5$ atoms in a 3-dimensional trap.
 Time $t=0$ corresponds to the beginning of the amplitude modulation; the modulation halts at $t=0.5$ms.
 (b) The band populations of the first and second bands as functions of time.}
 \label{fig:2ndp}
\end{figure}

The single-particle quasimomentum density distribution is given by the projection of the wave function onto the Bloch states, namely
\begin{equation}
|\Psi_n(q;t)|^2= \sum_{j=1}^N  |\langle \phi_q^n(x) | \psi_j(x;t) \rangle|^2,
\label{eq:density}
\end{equation}
thus the band population corresponds to $B_n(t)=\sum_q |\Psi_n(q;t)|^2$.
The populations of atoms of the first and second bands are shown in Fig.~\ref{fig:2ndp}; the numbers of atoms are complementary, and the sum remains nearly constant.
More precisely, the number of atoms in the ground band fluctuates on the order of 10$^{-3}$ due to the coupling between ground and 1st band. 
This is too small to affect the following discussion.

The hole dynamics in the ground band is almost decoupled from the  higher bands and was discussed in \cite{Heinze} both from semiclassical point of view and via numerical experiments.  A somewhat unsatisfactory aspect that remained is that the results of the semiclassical and numerical dynamics deviate from the experimental data on longer timescales. 
As we show below, semiclassical wave packets can be constructed more accurately, allowing us to observe some new features of the dynamics.

\begin{figure}[htbp]
 \begin{center}
 \includegraphics[width=9cm]{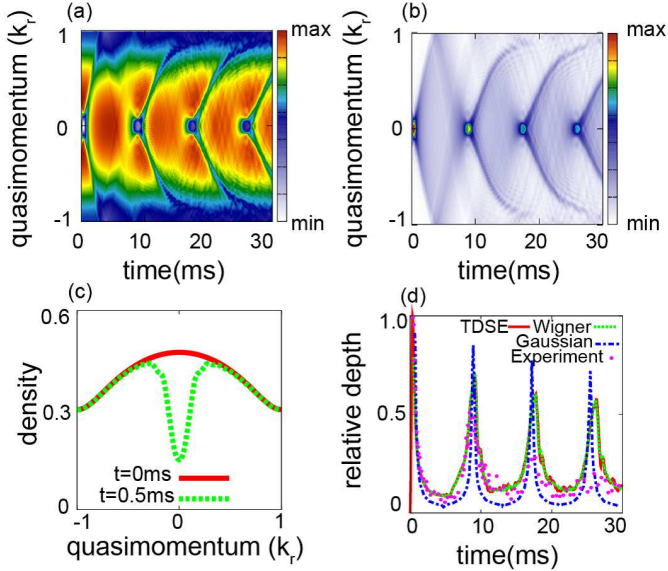}
 \end{center}
 \caption{(a) Time evolution of quasimomentum distribution of the ground band.
 (b) Differential quasimomentum distribution $h(q;t)$ of the ground band as a function of time.
 (c) Quasimomentum distributions at $t=0$ms (red solid line) and $t=0.5$ms (green dashed line) in the ground band.
 (d) The relative hole depth $d(t)$ as a function of time according to the full quantum evolution (`TDSE'), semiclassical TWA evolution (`Wigner'), experimental data for $N=81$, $s=2$ (`Experiment') and previous semiclassical approach (`Gaussian').}
 \label{fig:comparison}
\end{figure}

In Figs.\ref{fig:comparison} (a),(b), we show the results of our numerical experiment corresponding to Figure SF1 in \cite{Heinze}.
Fig.\ref{fig:comparison} (a) displays time evolution of quasimomentum distribution of the ground band and (b) shows differential distribution $h(q;t)=|\Psi_0(q;t=0)|^2-|\Psi_0(q;t)|^2$.
As we discussed above, the amplitude modulation with $E_\omega$=4.223 makes a single hole at the end of the pulse(Fig.\ref{fig:comparison} (c)).
After the modulation, the hole density spreads out in the ground band and comes back to $q=0$ periodically as seen in Fig.\ref{fig:comparison} (b).
The experimental paper focused on this revival feature and analyzed it with a semiclassical approach.
In Fig.\ref{fig:comparison} (d), we show a relative hole depth $d(t)=h(q=0;t)/{\rm max}\{h(q=0;t)\}$ defined at $q=0$ as a function of time and compare it to the experimental result.
Our TDSE approach shows good agreement with the experimental result in \cite{Heinze}.

\begin{figure}[htbp]
 \begin{center}
 \includegraphics[width=8cm]{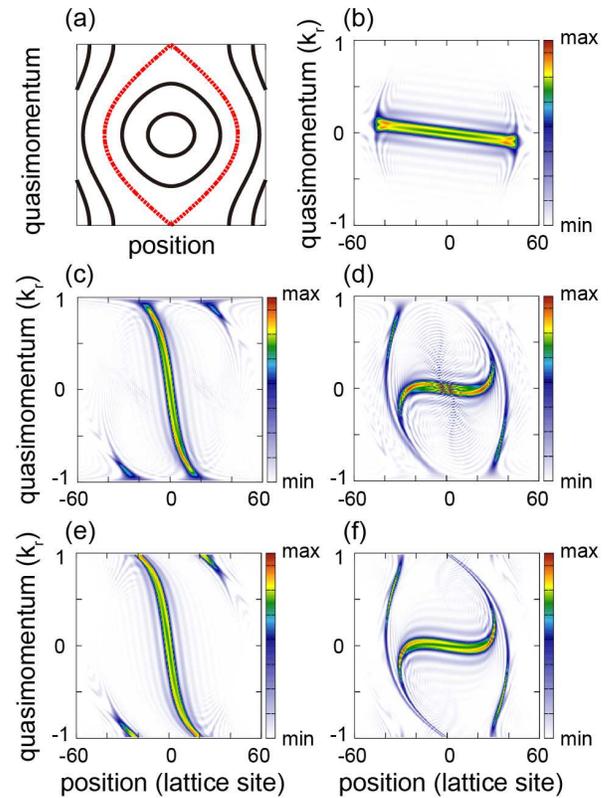}
 \end{center}
 \caption{(a) A conceptual figure of the phase space representation in the ground band.
 Each line shows isoenergy contours.
 Red dashed one represents a separatrix of the classical Hamiltonian $H_c$.
 Non-harmonic dispersion given by $H_c$ leads to dephasing as manifested in the distortion of the density distribution through phase space. The distortion is particularly noticeable toward the separatrix. See text for details.
  Differential Wigner distribution $\Delta\rho_{wig}(t)=\rho(t=0)-\rho(t)$ (hole distribution) of the ground band population at (b) $t=$0.5ms, (c) 4ms, and (d) 9ms.
 Semi-classical counterparts to the differential Wigner distributions of the first band component at (e) $t=$4ms and (f) 9ms.
 Note the time evolution of a classical function $f(x,q,t)$ is given by
 $\frac{d}{dt}f(x,q,t)=\frac{\partial f}{\partial q}\frac{\partial H_c}{\partial x}-\frac{\partial f}{\partial x}\frac{\partial H_c}{\partial q}$.
 }
 \label{fig:wigs}
\end{figure}

To analyze the dynamics of the hole in detail, let us recall the semiclassical approach used in Ref.\cite{Heinze}.
A classical Hamiltonian for a  single particle in the ground band is approximately given by $H_c=-J \cos(\pi q) + \nu x^2$ where $J$ is the tunneling parameter for the ground band.
$H_c$ has the same form as that of the nonlinear pendulum(Fig.~\ref{fig:wigs}(a).), with the role of the coordinate and momentum exchanged \cite{Kolovsky,Heinze}.  
The dynamics can be divided into two regimes, dipole oscillations corresponding to motion inside the separatrix on the phase portrait, and Bloch oscillations corresponding to motion outside the separatrix.
The dipole oscillations contribute to the hole revival, however $H_c$ does not lead to an isochronous motion, so that perfect hole revivals are prevented.
The initial distribution of fermions is approximated by an ensemble of classical phase points
enclosed by Fermi energy. Quantum observables are obtained by averaging over
this classical ensemble.  As a result of the amplitude modulation of the lattice,  a  hole is formed in the ground band. Ref.~\cite{Heinze} approximated the hole  by the gaussian ansatz. 
 This amounts to employing an effective single-particle description of the hole wave packet, that is the wave packet constructed of single-particle quasimomentum
 eigenstates is made gaussian with respect to quasimomentum, and the corresponding classical distribution is the Wigner transform of this wave packet.  
 The approach can be justified in the case of weak perturbations where the number of atoms excited out of the ground band is small. However, the following procedure provides 
 a more consistent  semiclassical distribution.  
Once $N$ single-particle eigenstates of the initial system are prepared,  TDSE generates  their evolution during the modulation pulse.
Then the corresponding $N$ Wigner distributions are calculated.  According to the prescription of the Truncated Wigner Approximation(TWA), one sums them up, and obtains quasiclassical `probability' distribution,
which can be time-propagated using the classical equations of motion.  
Such approach produces results that are almost indistinguishable from the full quantum evolution over tens of milliseconds.
The dynamics of Wigner function according to quantum and classical evolution is shown in Fig.\ref{fig:wigs}.

\begin{figure}[htbp]
 \begin{center}
 \includegraphics[width=8cm]{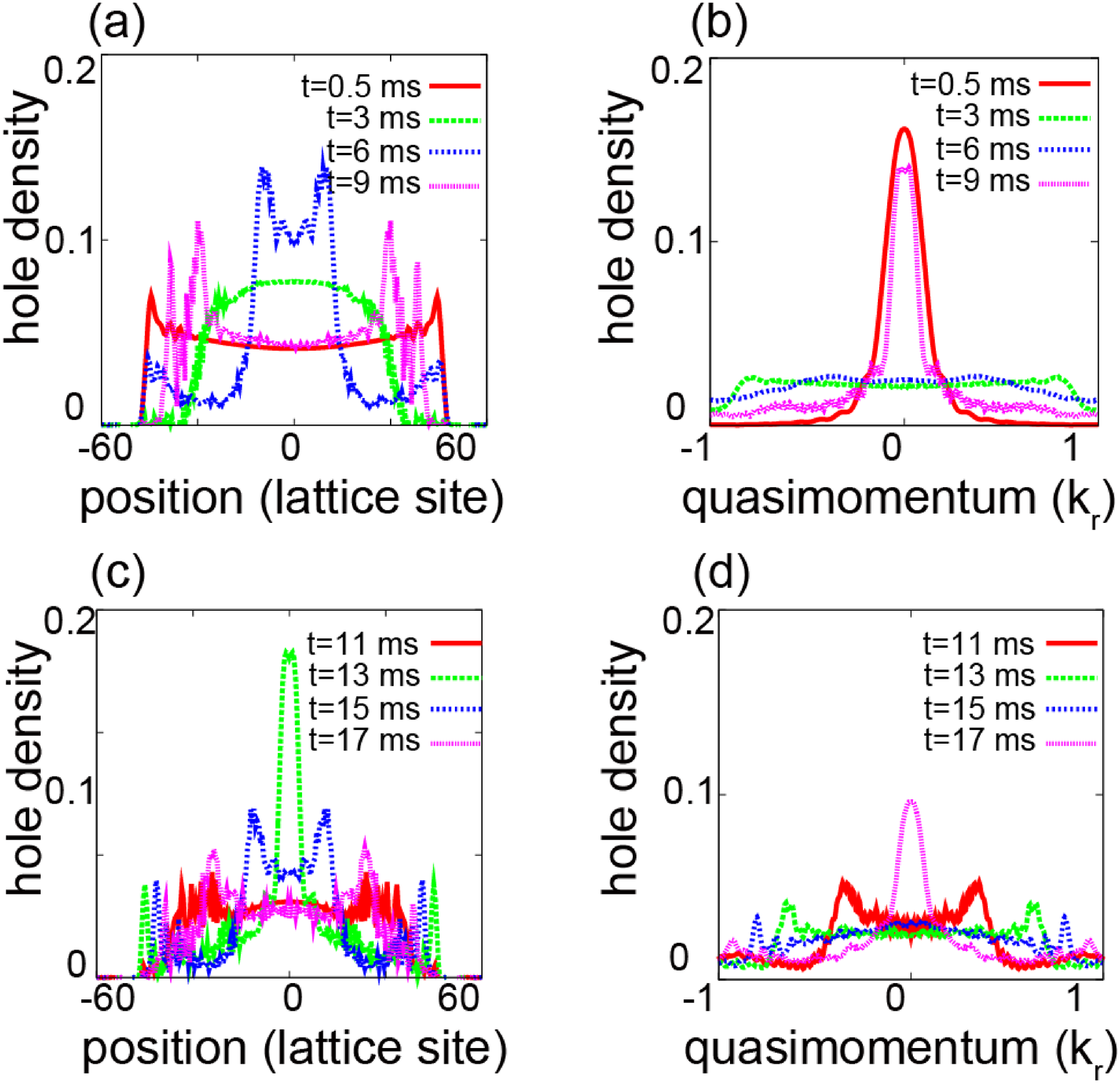}
 \end{center}
 \caption{Hole density represented in position space can be obtained by integrating the Wigner distribution over the quasimomentum coordinate. Likewise for quasimomentum representation.
Hole density shown at $t=$0.5, 3, 6 and 9ms  (a) in position space, and
 (b) in quasimomentum space.
 Hole density in the quasimomentum space behaves like gaussian up to 9ms.
 However, after that the density behaves as if consisted of two traveling wave packets.
 Hole density at $t=$11, 13, 15 and 17ms shown (c) in position space and
 (d) in quasimomentum space.
 }
 \label{fig:densities}
\end{figure}

We also show the time evolution of the hole density in the position and quasimomentum space in Fig.\ref{fig:densities}.
These figures clearly show that an initial density profile in the position space is more rectanguler than gaussian. 
And this makes the hole behave like a pair of symmetrically propagating wave packets in phase space. 
As easily seen, the present numerical implementation reproduces experimental data better than the previous semiclassical approach of \cite{Heinze}. 
There is asymmetry in the form of hole depth oscillations, namely pronounced irregularity near its minima, which is correctly reproduced by the new approach(See Fig.\ref{fig:comparison} (c)). 
However, the new approach cannot reproduce the observed rate of decline of the peaks.
This means the hole has long coherence time at temperature $T=0$.

One of the purposes of the present paper is to address and investigate the unresolved discrepancy, pertaining to the decline of the pulse peaks, between the model and experimental dynamics. 
We shall check a few possible causes: dimensionality, finite temperature and imperfection of the trap.
There might be other conceivable causes, {\it e.g.} small p-wave interaction among fermions and dissipation due to the optical lattice potential, but they are outside the scope of this paper.
At any rate, the origin of the discrepancy is all that remains unexplained. 

\section{Discussions: Dephasing of the hole due to dimensionality, finite temperature, and imperfect trap} 

\subsection{Finite temperature effects}
We include thermal effects by  preparing a corresponding initial state mixture. During and after the modulation pulse the system is 
assumed to be decoupled from the environment, {\it i.e.} all thermal effects in the present consideration are only due to the initial distribution.

\begin{figure}[htbp]
 \begin{center}
 \includegraphics[width=7cm]{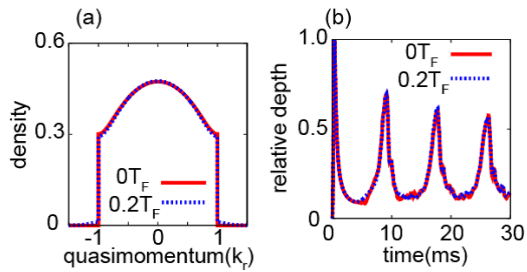}
 \end{center}
 \caption{(a) Initial quasimomentum distributions for the temperature $T=0$ (red solid line) and 0.2$T_F$ (blue dashed line), respectively.
 (b) The relative hole depth defined as for Fig.~\ref{fig:comparison}(d) at the corresponding temperature.}
 \label{fig:temp-1d}
\end{figure}

The single-particle density distribution is represented as
\begin{equation}
|\Psi(x;t)|^2= \sum_{j} |\psi_j (x;t)|^2 \frac{1}{e^{\beta(\epsilon_j-\mu)}+1}
\label{eq:1d-tdwf}
\end{equation}
where $\beta=E_r/k_B T$, $\mu$ and $\epsilon_j$ are scaled inverse temperature, scaled chemical potential, and eigenenergies of the unperturbed {\it i.e.}~unmodulated system, respectively.
The scaled chemical potential is calculated by solving $N-\sum_j \frac{1}{e^{\beta(\epsilon_j-\mu)}+1}=0$.
To compare our results with the experimental ones, we employ the Fermi temperature $T_F=\epsilon_{F}/k_B$ as the unit of temperature, where $\epsilon_{F}$ is the Fermi energy.
The results shown in Fig.~\ref{fig:temp-1d} for this case reveal modest difference from $T=0$.

\begin{figure}[htbp]
 \begin{center}
 \includegraphics[width=9cm]{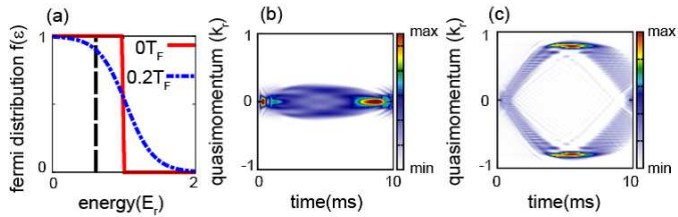}
 \end{center}
 \caption{
 (a) Fermi distribution $f(\epsilon)$ as a function of energy with number of atoms $N=81$.
 Red solid and blue dashed lines show the results with $T=0$ and $0.2T_F$, respectively.
 The separatrix in the energy domain $E_s=0.59$ also shown by a black dashed line.
 Subtracted quasimomentum distributions with (b)$j=1-5$ and (c)$j=40-45$ are shown within $t=$10ms.}
 \label{fig:temp-effect}
\end{figure}

To distinguish different roles played by states near the bottom of the ground band $q \sim 0$ and those near the band edge $q \sim \pm1$, we follow the dynamics of the low and high energy eigenstates separately, retaining only those states that reside inside the separatrix.
Figs.~\ref{fig:temp-effect} (b) and (c) show the subtracted quasimomentum density distributions constructed for limited ranges of index $j=1-5$ and $j=40-45$, respectively.
The low energy range ($j=1-5$) mainly consists of Bloch states $\phi_q$ with $q\sim 0$, thus representing the hole state localized near $q \sim 0$.
The hole revival near $q \sim \pm1$ seemingly lags behind because of the location
close to the separatrix where the trajectories evolve very slowly (see Fig.\ref{fig:wigs} (c)). In this way, the revival time of the trajectories vary over a range depending on whether they evolve near $q\sim 0$ or near $q\sim \pm1$.
The time lag between low and high energy trajectories
 could be clearly seen in Fig.~\ref{fig:temp-effect}(c).
This lag makes the revival tail off so that the hole peaks become asymmetric
 in hole depth $d(t)$ (Fig.\ref{fig:temp-1d}(b),$T=0$).
Even in the case of finite temperatures, these features are not changed because the thermal excitations merely reduce the small number of particles in the high energy region around the separatrix. 
This effect smoothes the $d(t)$ curve, making the asymmetry less conspicuous.
Indeed, Fig.~\ref{fig:temp-effect}(a) shows the 1D Fermi distribution with T=0 and $0.2 T_F$, the latter clearly indicating the reduced contribution from the high energy region due to thermal excitation.

\subsection{Three dimensional effects}
\begin{figure}[htbp]
 \begin{center}
 \includegraphics[width=7cm]{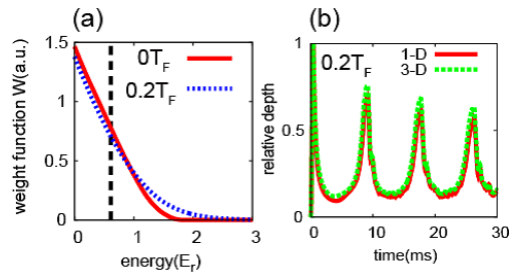}
 \end{center}
 \caption{(a) Weight function $W$ as a function of energy $\epsilon_{j_x}$ with $T=0$(red solid line) and $0.2T_F$(blue dashed line).
 The separatrix in the energy domain also shown by a black dashed line same as Fig.\ref{fig:temp-effect}.
 (b) Relative depth with temperature $T=0.2T_F$.
 Red solid and green dashed curves correspond to the 1-dimensional and 3-dimensional results, respectively.}
 \label{fig:temp-3d}
\end{figure}

Including the 3D effect requires the density of states arising from the other two degrees of freedom, ignored up to now. 
We  extend the 1D finite temperature representation to 3D with eigenfunctions $\psi_{j_x}$, $\psi_{j_y}$ and $\psi_{j_z}$, and eigenenergies $\epsilon^x_{j_x}$, $\epsilon^y_{j_y}$ and $\epsilon^z_{j_z}$ for each direction,
\bea
|\Psi(x,y,z;t)|^2 &=& \sum_{{j_x},{j_y},{j_z}} |\psi_{j_x} (x;t) \psi_{j_y} (y;t) \psi_{j_z} (z;t)|^2   \nonumber\\
& \times & \frac{1}{e^{\beta(\epsilon_{{j_x},{j_y},{j_z}}-\mu_{3d})}+1}
\label{eq:3d-tdwf}
\eea
where the total energy is $\epsilon_{{j_x},{j_y},{j_z}}=\epsilon^x_{j_x}+\epsilon^y_{j_y}+\epsilon^z_{j_z}$, and the 3D scaled chemical potential is calculated by solving $N_{3d}- \sum_{{j_x},{j_y},{j_z}} \frac{1}{e^{\beta(\epsilon_{{j_x},{j_y},{j_z}}-\mu_{3d})}+1}=0$.
Here, we assume  for simplicity that the trap is an isotropic 3D parabolic lattice and the amplitude modulation is applied only to the x-direction.
Therefore, the density reduced distribution is given by
\begin{equation}
|\Psi(x;t)|^2= \sum_{{j_x}} W(j_x) |\psi_{j_x} (x;t)|^2,
\label{eq:reduced-density}
\end{equation}
where the weight function $W(j_x)=\sum_{{j_y},{j_z}} \frac{1}{e^{\beta(\epsilon_{{j_x},{j_y},{j_z}}-\mu)}+1}$.
We set the number of atoms $N_{3d}$ to $5.3\times 10^5$.
The weight functions with temperature $T=0$ and $0.2T_F$ are shown in Fig.~\ref{fig:temp-3d}(a).
The functions resemble the 1D Fermi distributions, therefore the discussions for the 1D case remain applicable.
Note that, in the case of $T=0$, the component of high energy region below the separatrix is suppressed.
Thus we observe much more robust hole-revival than in the 1D case.
However, the results shown in Fig.~\ref{fig:temp-3d}(b) with $T=0.2T_F$ are not dramatically different from the 1D case.
No dramatic change is thus reproduced by any of these effects arising from the density distribution.

\subsection{Imperfect trap}

\begin{figure}[htbp]
 \begin{center}
 \includegraphics[width=8cm]{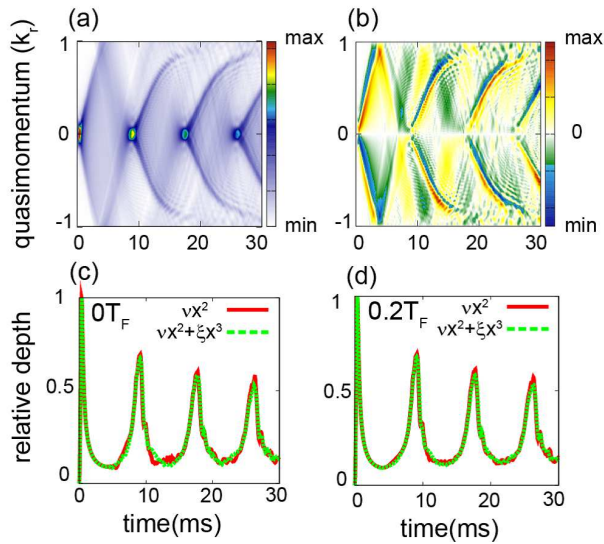}
 \end{center}
 \caption{(a) Subtracted quasimomentum distribution with cubic potential $\xi=5.0\times 10^{-8}$.
 (b) Asymmetry of the quasimomentum distribution $a(t,q)$.
 Hole depth $d(t)$ at (c) absolute zero temperature $T=0$ and (d) finite temperature $T=0.2T_F$.}
 \label{fig:cube}
\end{figure}

We assume the same degree of imperfection may occur in the Hamburg experiment as in the Aarhus experiment\cite{Arlt}.
In adding an extra cubic potential term $\xi x^3$ to the Hamiltonian $H_0$,
we set the cubic parameter $\xi=5\times 10^{-8}$ slightly above the experimental value of $\xi \sim 2\times 10^{-8}$ measured in the Aarhus experiment to see how a deformation of the trap potential affects the dynamics of the ground band atoms.
Note that, if the cubic potential is used without restriction, it ends up allowing a loss of atoms due to tunneling. 
However, we diagonalize the Hamiltonian defined in the limited position space $|x| < \nu/\xi$ so that the potential term is always positive. 
This insures that the number of atoms is conserved in our numerical simulation.
Here we use a 1D system as given by Eq.~(\ref{eq:1d-tdwf}).
The results shown in Fig.~\ref{fig:cube}(a) indicate little difference from the perfectly parabolic trap. 
Although no fundamental difference shows up at $q=0$, we observe some asymmetry induced by the cubic term in the side wings in (b).
Here we define  $a(t,q)=|\Psi(t,q)|^2-|\Psi(t,-q)|^2$ as a measure of asymmetry.
Figs.~\ref{fig:cube}(c) and (d) show the hole depth $d(t)$ at temperatures $T=0$ and $0.2 T_F$, respectively.
We do not observe any qualitative differences stemming from the cubic parameter because it modifies neither the phase space structure nor the separatrix in the ground band strongly.
However, the cubic term $\xi=5\times 10^{-8}$ modifies the phase space structure in the higher bands, and totally breaks symmetric features.
Since no such asymmetric features were observed in the Hamburg experiment\cite{Heinze}, the cubic parameter would not be bigger than $\xi=5\times 10^{-8}$ in the actual experiment.

\section{Conclusions}
In this paper, we reexamined the hole dynamics of a fermionic system in an amplitude modulated parabolic lattice
using  numerical calculations as well as the refined semiclassical approach.
It is remarkable that careful preparation of the semiclassical distribution  allows to reproduce the exact quantum dynamics on long timescales using the same classical Hamiltonian as in \cite{Heinze}. 
Several qualitative features of the dynamics are successfully reproduced. At the same time, decay of the hole wave packet in the experiment happens
faster than in our numerical calculations. To clarify what affects the decay of the hole state, we checked three possible causes: finite initial temperature of the system, three-dimensionality of the setup, and anharmonic distortion of the parabolic trap. None of them can explain the remaining discrepancy. 
We also checked two other causes possibly attributable to the imperfection of the amplitude modulation, namely a mixture of the second harmonics, and a white noise in the excitation frequency. We find that they do not substantially alter the hole dynamics.
The discrepancy from the experiment may be caused by loss of atoms or heating by environment during dynamical evolution of the atoms after the pulse.
A closer experimental examination might be useful.
Numerical results show the hole dynamics has an intrinsically long coherence time in the limit of a perfectly isolated lattice system.
It leads to a suggestion that the hole state could be used as a quantum element for an interferometer, as its bosonic counterpart of BEC.
Thus the manipulations of the hole state subject to a two-color optical lattice\cite{Sebas} is a possible future extension of this work.  Another natural extension is a study of interacting mixture of fermionic atoms\cite{Heinze}, and dynamics of nonlinear waves\cite{Protopopov} in such systems.

\acknowledgements
This work was partly supported by JSPS KAKENHI Grant
No. 26400416. T.Y. acknowledges support from the JSPS
Institutional Program for Young Researcher Overseas Visits. 
A.P.I thanks his Hamburg collaborators C. Becker, J.Heinze, and K.Sengstock for insightful discussions and explanations, and acknowledges support from NWO and University of Electro-Communications.

\end{document}